\documentclass[conference]{IEEEtran}
\IEEEoverridecommandlockouts

\usepackage{amsmath,amsfonts,amssymb, cite}
\usepackage{caption}
\usepackage{graphicx}
\usepackage{booktabs}
\usepackage[colorlinks=true, allcolors=blue]{hyperref}
\usepackage{authblk}

\title{Federated Learning for Distributed Spectrum Sensing in NextG Communication Networks} 

\author[a]{Yi Shi}
\author[b]{Yalin E. Sagduyu}
\author[b]{Tugba Erpek}
\affil[a]{Commonwealth Cyber Initiative, Virginia Tech, Arlington, VA, USA}
\affil[b]{National Security Institute, Virginia Tech, Arlington, VA, USA}

\begin{document} 
\maketitle

\begin{abstract}
NextG networks are intended to provide the flexibility of sharing the spectrum with incumbent users and support various spectrum monitoring tasks such as anomaly detection, fault diagnostics, user equipment identification, and authentication. For that purpose, a network of wireless sensors is needed to monitor the spectrum for signal transmissions of interest over a large deployment area. Each sensor receives signals under a specific channel condition depending on its location and trains an individual model of a deep neural network accordingly to classify signals. To improve the accuracy, individual sensors may exchange sensing data or sensor results with each other or with a fusion center (such as in cooperative spectrum sensing). 

In this paper, distributed federated learning over a multi-hop wireless network is considered to collectively train a deep neural network for signal identification. In distributed federated learning, each sensor broadcasts its trained model to its neighbors, collects the deep neural network models from its neighbors, and aggregates them to initialize its own model for the next round of training. Without exchanging any spectrum data, this process is repeated over time such that a common deep neural network is built across the network while preserving the privacy associated with signals collected at different locations. Signal classification accuracy and convergence time are evaluated for different network topologies (including line, star, ring, grid, and random networks) and packet loss events. In addition, the reduction of communication overhead and energy consumption is considered with random participation of sensors in model updates. The results show the feasibility of extending cooperative spectrum sensing over a general multi-hop wireless network through federated learning and indicate the robustness of federated learning to wireless network effects, thereby sustaining high accuracy with low communication overhead and energy consumption.  
\end{abstract}


\section{INTRODUCTION}
\label{sec:intro}  

Federated learning allows multiple clients to collaborate with each other to train a machine learning model. In the typical setting of federated learning, a (centralized) server orchestrates the learning process \cite{McMahan17:FL}. Each client trains its individual model with its local data. Then, clients send their models to the server. The server aggregates these models and sends the aggregate model back to clients. Then, each client initializes its model with this aggregate model for the next round of update. This process is repeated over multiple rounds until a convergence criterion is met. 
%
As clients send only their local models and do not share their local data, federated learning offers privacy for the learning process. Overall, the potential benefits of federated learning include (i) low data processing requirement for each client, (ii) high accuracy of global model even if individual clients may sustain low accuracy without federated learning, (iii) privacy (since no local data is shared), and (iv) communication efficiency (since the model parameters to be shared is smaller than the training data) \cite{Survey0:FL, Survey1:FL, Survey2:FL, Survey3:FL}.

Recently, federated learning has been considered for wireless communication systems \cite{Wireless5:FL, Wireless4:FL, Wireless1:FL, Wireless2:FL, Wireless3:FL}. Examples include mobile edge networks \cite{MEC:FL}, Internet of Things (IoT) \cite{FL:IoT}, 5G \cite{5G:FL}, and 6G \cite{6G:FL}. Typically, the client-server paradigm is executed over single-hop wireless links in these examples. There have been also extensions of federated learning to more general network settings such as social networks \cite{Multi1:FL} and general graph models \cite{Multi2:FL}.   
\begin{figure}[h!]
\centering
   \includegraphics[width=0.75\columnwidth]{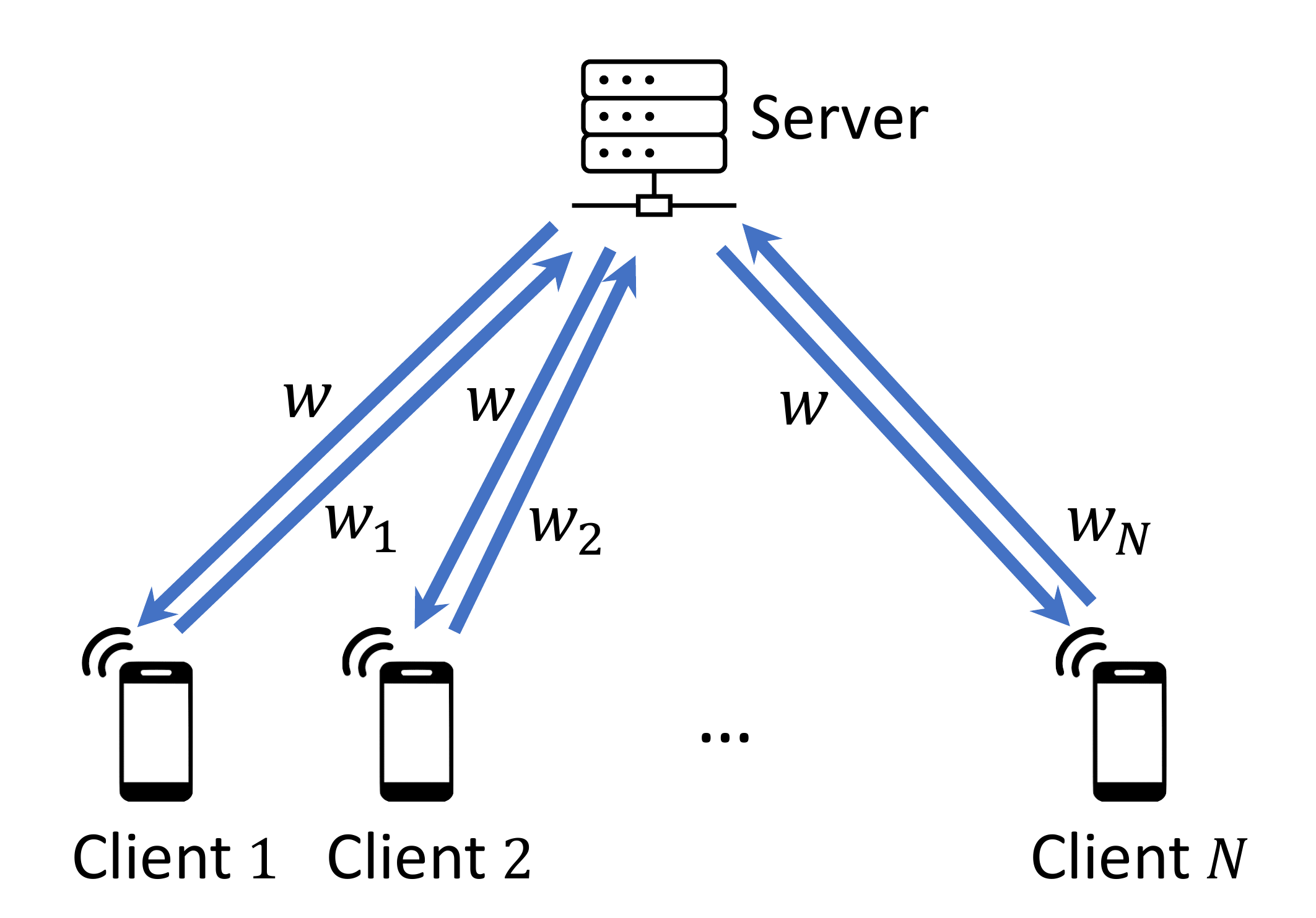}
   \caption{Client-server model of federated learning (client $i$'s local model is denoted by $w_i$ and global model is denoted by $w$).}
\label{fig:FL} 
\end{figure}

The typical client-server model of federated learning (illustrated in Fig.~\ref{fig:FL}) requires a centralized server to process all local models and broadcast the global model updates to clients. While this information exchange can be executed over wireless links, in addition to the need of extra equipment, this centralized scheme raises the vulnerability of a single sensor failure, i.e., if the server is hacked or jammed \cite{Attack:FL}, the entire federated learning system cannot function properly. In addition, it is possible that some of clients are not in the communication range of the server. In this paper, we study a distributed federated learning scheme that does not require a centralized server, can be executed over a multi-hop wireless network, 
and thus is suitable for wireless network applications such as distributed spectrum sensing. We assume that each sensor in the network collects its own data and does not need to exchange data with other sensors. The local data is used to train/update a model at each sensor, which may not work well for a general dataset including data from all sensors. Thus, each sensor broadcasts its trained model to neighbors within its transmission range. After receiving models from neighbors, each sensor computes a new model using federated averaging on its own model and neighbors' models. This new model can provide better performance than each sensor's own model as it incorporates data from all sensors. This process can be repeated many rounds until the models at sensors converge.

\begin{figure}
  \centering
   \includegraphics[width=1\columnwidth]{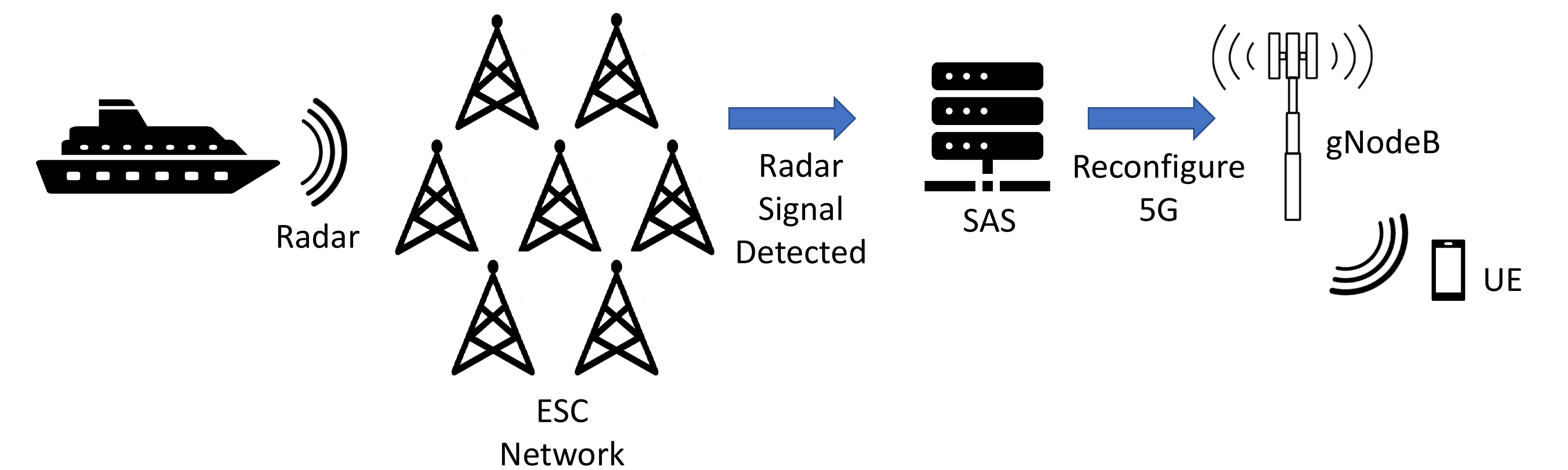}
   \caption{The wireless sensor network for ESC operating in the CBRS band (the ESC network detects radar signals as incumbent users and informs the SAS to reconfigure the 5G communications and prevent interference with radar signals).}
\label{fig:CBRS} 
\end{figure}

We apply the distributed federated learning scheme to the spectrum sensing problem. One real-world application scenario is the Environment Sensing Capability (ESC) for 3.5GHz Citizens Broadband Radio Service (CBRS) \cite{FCC}, where the radar and 5G communications  need to co-exist in the same spectrum band. The ESC sensors need to detect the incumbent (radar) signal such that the Spectrum Access System (SAS) can configure the 5G communications system to avoid interference with the incumbent user (as illustrated in Fig.~\ref{fig:CBRS}). That is, sensors need to identify whether a sensed signal belongs to a target signal or not. Each sensor can sense the spectrum, collect raw I/Q data, use a deep learning model to identify whether a target signal is present \cite{DL:CBRS}. 

If such a model is built by a single sensor's data, this model may not work well in general, i.e., the accuracy will drop if the data is collected by another sensor. In addition, a single sensor may only monitor a limited area. Thus, we consider distributed spectrum monitoring with a multi-hop wireless sensor network, where each sensor broadcasts its model to neighbors within its communication range and receives neighbors' models, as well. Then, a federated averaging technique is applied to the received models and a sensor's own model to obtain an aggregate model. This process needs to be repeated multiple rounds so that all sensors' models can be propagated across the entire network and each sensor's model can be improved to converge to a highly accurate model.

We apply federated learning in different multi-hop network topologies, including line, star, ring, and grid networks as well as random networks. For any topology, distributed federated learning needs more rounds to converge than the conventional federated learning scheme executed over single-hop links, such as considered in previous work \cite{Attack:FL}. In general, federated learning can converge faster when applied over a topology with more connections. The star topology looks the same as the standard topology used for conventional federated learning, i.e., the server is connected with all clients. However, the learning processes under conventional and distributed federated learning are different. As a consequence, federated learning in the star topology has the worst accuracy among all topologies considered in this paper. The performance of federated learning in the grid topology is the best one since most sensors have four neighbors. We also find that distributed federated learning is robust to packet losses. There is no drop of accuracy even when we impose $50$\% packet loss. We further consider the scheme that sensors do not participate in all rounds of updates. Instead, each sensor decides to broadcast its model with a certain probability. The benefit is that we can reduce the communication overhead and (transmit) energy consumption at sensors. In this setting, distributed federated learning maintains high accuracy when the broadcast probability is $25$\% and the accuracy remains acceptable ($88.65$\%) even when the broadcast probability is reduced to $10$\%. 

The remainder of the paper is organized as follows. Section~\ref{sec:sensing} presents the distributed operation of federated learning over a multi-hop wireless network. Section~\ref{sec:perf} evaluates the performance of distributed federated learning for spectrum monitoring under different topologies and packet loss events, and discusses how to reduce the communication overhead and energy consumption.  Section~\ref{sec:conclusion} concludes the paper.

\section{Distributed Spectrum Sensing}
\label{sec:sensing}

\subsection{Problem Description and a Centralized Solution}

Spectrum sensing is a fundamental task for wireless communications and has important applications, including dynamic spectrum access (where signals of primary (incumbent) users need to be identified), anomaly detection, and user authentication.  In this paper, we consider a spectrum sensing problem that aims to determine whether a target signal appears or not by using sensing results from multiple sensors. We consider a network of sensors that collects I/Q data. There is one 
transmitter that generates either a target signal or some other signal(s). The sensors need to determine whether the target signal appears. For wireless networks with complex channel dynamics, it has been a common practice to use deep learning for wireless signal classification \cite{DL:wireless1, DL:wireless2}. Each sensor builds its own deep learning model to classify sensing results as either with target signal or without target signal. 
To build a general model that can be used to classify data collected by any sensors, we need to have training data from all sensors. Thus, we can have a centralized solution by letting sensors transmit their data to one sensor (acting as a fusion center) and this sensor trains a model based on data from all sensors. This approach has high communication overhead to transmit data and high computation burden at one sensor. Moreover, there may be privacy concerns due to transmitting all the data samples to one sensor and storing them at this sensor.

\subsection{Distributed Federated Learning}

The conventional federated learning scheme can reduce the communication overhead and the privacy concern by sending the trained models instead of data. For that purpose, a client-server structure is used. Initially, each client starts with a random model. Then, federated learning has the following steps in each round of update.
\begin{enumerate}
    \item Based on the current model and local data, each client trains and updates its model, then sends its model to the server.
    \item The server calculates a global model by applying federated averaging on received client models and broadcasts this global model to all clients. 
    \item Each client updates its model by the global model (i.e., it initializes its model by the global model and retrains it with its own data).
\end{enumerate}
The learning process ends when the global model converges. This scheme cannot be applied for distributed spectrum sensing because it is not practical to assume that there is a centralized server. One may want to specify a sensor as the server. However, there may not exist a sensor that can communicate with all other sensors  over a single hop (i.e., multi-hop network operation is needed).

\begin{figure}
  \centering
  		\includegraphics[width=1\columnwidth]{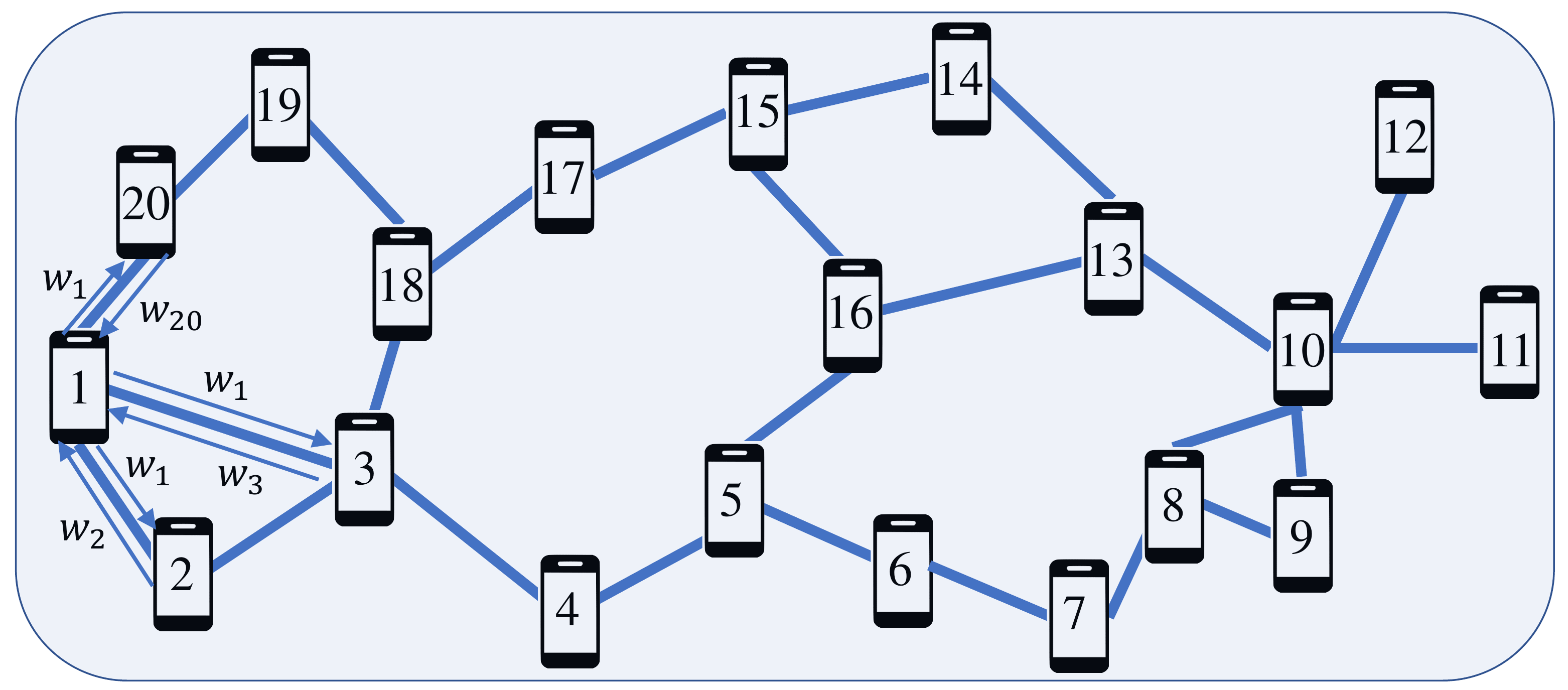}
  		\caption{Distributed federated learning over a multi-hop network (e.g., sensor 1 broadcasts its model $w_1$ to its neighbor sensors 2, 3, and 20 and receives theirs models $w_2$, $w_3$, and $w_{20}$ in one instance of the random multi-hop network).}
		\label{fig:general}
		
\end{figure}

To solve the distributed spectrum sensing problem, we consider a distributed federated learning scheme by removing the need of a server in the conventional federated learning scheme. It has the following steps in each round of update.
\begin{enumerate}
    \item Based on the current model and local data, each sensor trains and updates its model.
    \item Each sensor sends the update to its neighbors and receives the update from its neighbors.
    \item Each sensor calculates a new model by applying federated averaging on the received neighbor models and its own model.
\end{enumerate}
The learning process ends when each sensor's model converges. We assume that all sensors are connected via multi-hop communications so that each sensor can receive model updates from all other sensors either directly (for neighbors) or indirectly (for non-neighbors). This assumption is needed to ensure that all sensor models will converge to the same model. Distributed federated learning over an instance of multi-hop wireless network is illustrated in Fig.~\ref{fig:general}.
A feedforward neural network (FNN) is used as the deep neural network model at each sensor. The FNN properties are shown in Table~\ref{table:FFN}. The input to the FNN model is the phase shift and the power level analyzed from the raw I/Q data. The weights of local models are different for all sensors. 
To simplify the description, we assume a synchronized approach that all sensors participate in one round of update and one sensor can receive all its neighbor models. This assumption can be removed. As long as remaining updates still guarantee sufficient number of updates from neighbors, the distributed federated learning scheme can still converge (see Section~\ref{subsec:skip}). 


\begin{table}
	\caption{FNN properties.}
	\vspace{0.25cm}
	\centering
	{\small
		\begin{tabular}{c|c}
		Input size & 32 \\ \hline
		Output layer size & 2 \\ \hline
Hidden layer sizes & $128, 64, 32$ \\ \hline
Dropout rate & $0.2$ \\ \hline
Activation function & Relu (hidden layer)\\ & Softmax (output layer) \\ \hline
Loss function & Crossentropy \\ \hline
Optimizer & RMSprop \\ \hline
Number of parameters & 14,626
		\end{tabular}
	}
	\label{table:FFN}
\end{table}


\section{Performance Evaluation}
\label{sec:perf}

To evaluate the performance of distributed federated learning for spectrum sensing, we consider a transmitter that can transmit either QPSK (target signal) and BPSK (other signal). We analyze phase shift and power level for raw I/Q data to get features (namely, the input to a deep learning model). We consider features for $16$ bits as one sample. For BPSK, there are $32$ features. For QPSK, we calculate two sets of phase shift and power level for $2$ bits so that the number of features in a sample is also $32$.
\begin{figure}[t]
  \centering
   \includegraphics[width=0.7\columnwidth]{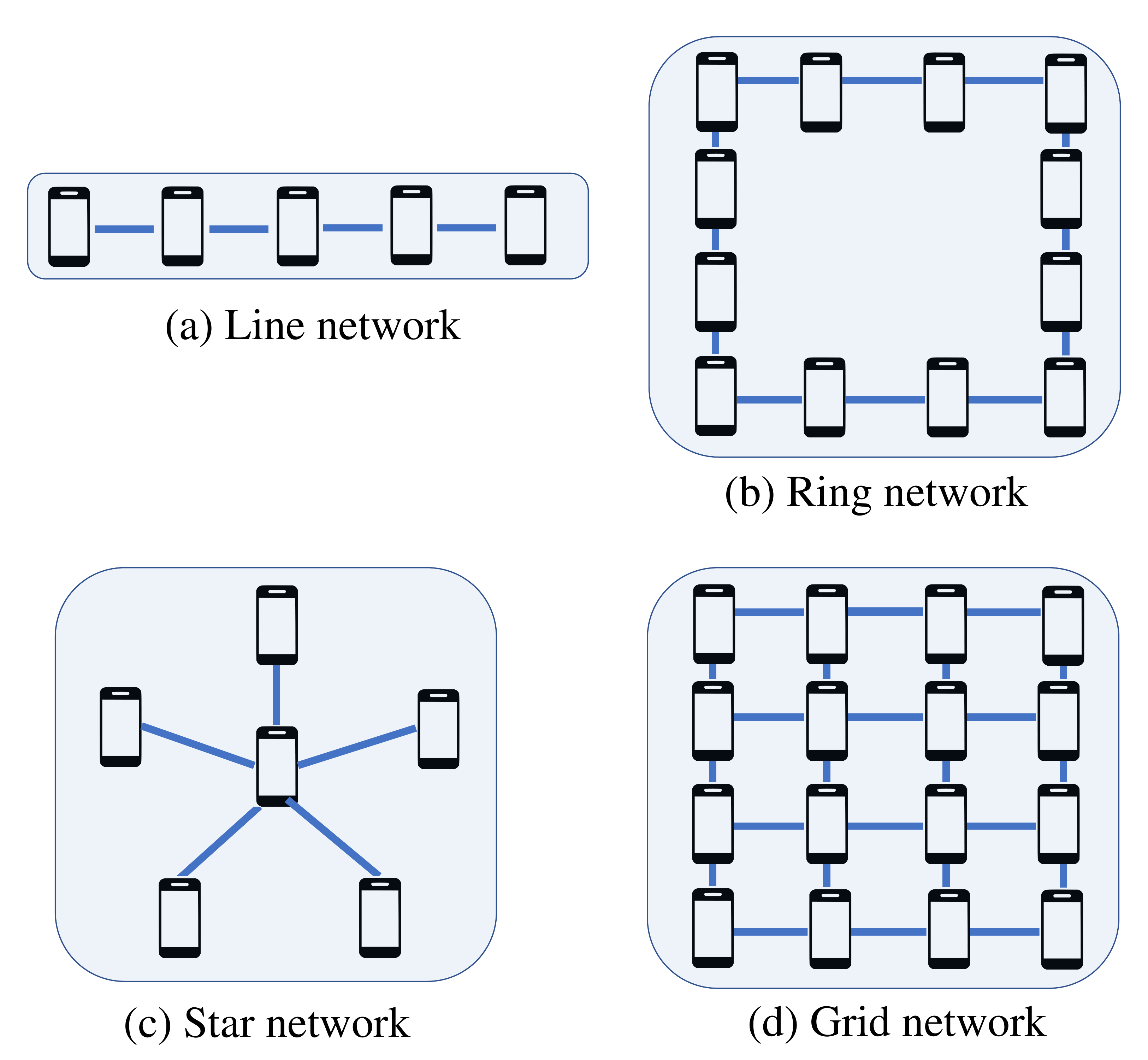}
   \caption{Line, ring, star, and grid network topologies.}
\label{fig:topologies} 
\end{figure}

The transmitter is deployed at $(0,0)$ while sensors are deployed in $[100,1000]\times[100,1000]$. The communication range is $400$. We consider different network topologies including line, ring, star, and grid networks (illustrated in Fig.~\ref{fig:topologies}), as well as a random network.

\begin{itemize}
    \item In the line topology, except two end sensors, all other sensors are connected with two neighbors. We assume that the line topology has five sensors at $(100,100), (300,300), (500,500), (700,700)$, $(900,900)$. 
    \item In the ring topology, all sensors are connected with two neighbors. We assume that the ring topology has $12$ sensors by removing $4$ sensors in the center from the grid topology.
    \item In the star topology, the central sensor is connected with all other sensors. We assume that the star topology has six sensors at $(500,500)$ and $(500 + 400 \cos{72 k},500 + 400 \sin{72 k})$ for $k \in \{0, 1, 2, 3, 4\}$. 
    \item In the grid topology, each sensor is connected with up to four neighbors at left, right, up and down. In particular, we assume that the grid topology has $16$ sensors at $(100+300i, 100+300j)$ for $i, j \in \{0,1,2,3\}$.
    \item In the randomly generated network, locations are randomly selected for $20$ sensors.
\end{itemize}

\begin{figure}[t]
  \centering
\includegraphics[width=\columnwidth]{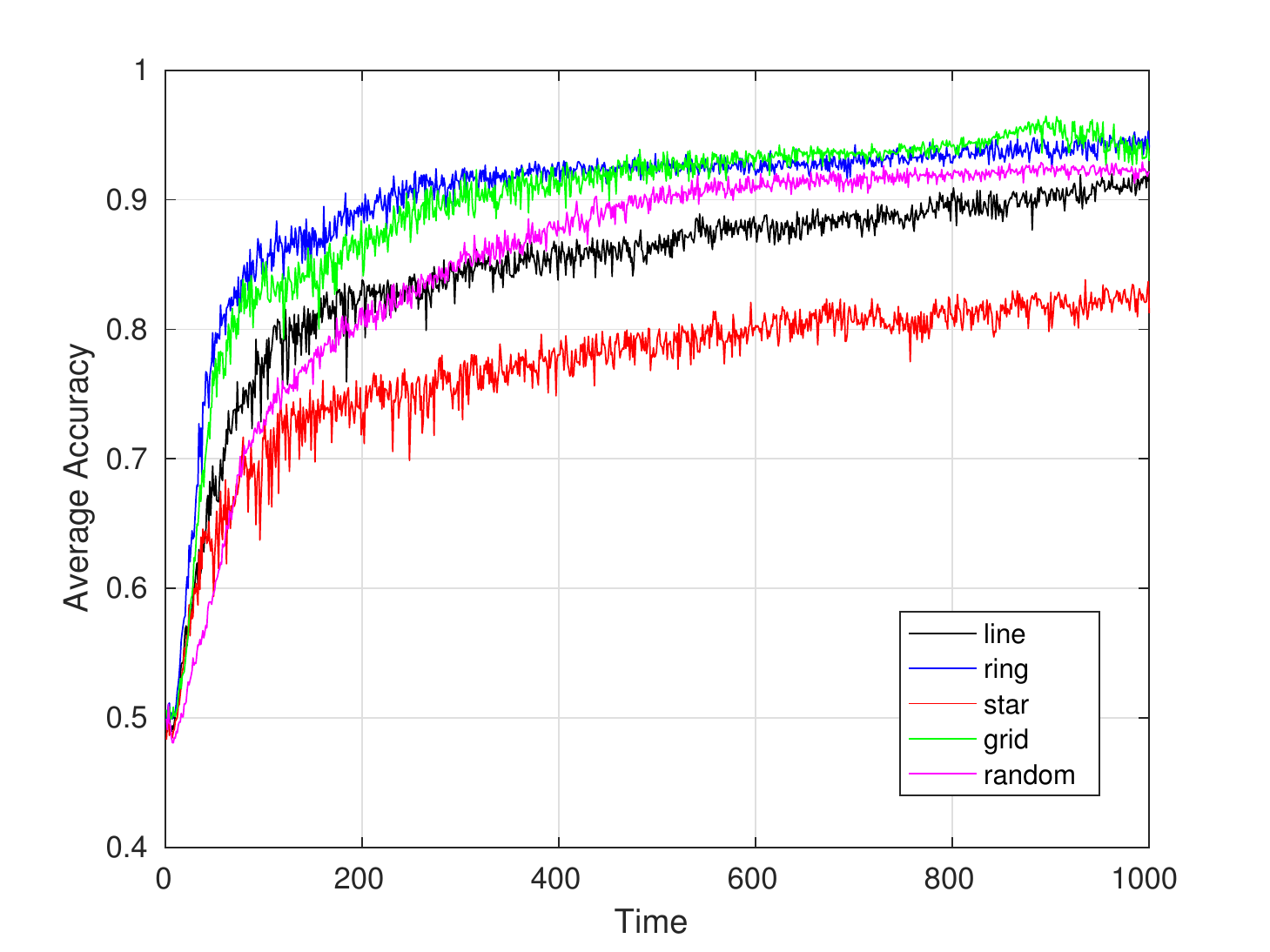}
\captionof{figure}{The average accuracy of federated learning achieved over time in different topologies.}
\label{fig:curve} 
\end{figure}


To measure the performance, we calculate the accuracy of each sensor model for a dataset of sensing data from all sensors. We use this test data to ensure that the trained models can achieve a good performance not for one sensor's data but for any sensor's data. For the case that there are sufficient number of sensors, we can claim that the trained model can work for a sensor placed at any location. Although such a dataset may not be available in practice, e.g., sensors do not want to share data due to privacy, we can still use it to evaluate the performance of distributed federated learning. To compare networks with different numbers of sensors in different topologies, we calculate the average accuracy (averaged over all sensors). We observe that this average accuracy fluctuates over rounds. Thus, we analyze when the best average accuracy converges. In particular, the best average accuracy in round $t$ is the maximum of average accuracies in rounds $1$ to $t$. We claim convergence if the best average accuracy does not increase by a small value $\varepsilon$ for continuous $M$ rounds. We set $\varepsilon = 0.01$ and $M = 100$ for numerical results.
\begin{table}[t!]
	\caption{Performance of federated learning for different network topologies.}
	\vspace{0.25cm}
	\centering
	{\small
		\begin{tabular}{c|c|c} 
Topology & Time & Best average accuracy \\ \hline \hline
Line & $900$ & $0.9098$ \\ \hline
Ring & $580$ & $0.9357$ \\ \hline
Star & $794$ & $0.8237$ \\ \hline
Grid & $581$ & $0.9389$ \\ \hline
Random & $704$ & $0.9221$ 
		\end{tabular}
	}
	\label{table:converge}
\end{table}
\subsection{Effects of Different Network Topologies}

Fig.~\ref{fig:curve} shows that the average accuracy of federated learning for each topology improves over time (rounds of updates). The average accuracy is about $0.5$ at the beginning since a random model is initialized. After $1000$ rounds, federated learning achieves average accuracy close to $0.9$ for all topologies, except the star topology. This seems to be counterintuitive since the star topology is similar to the client-server structure in conventional federated learning. However, the different update processes affect the performance. In conventional federated learning, the server collects all client models and performs federated averaging to obtain the global model, and clients use the global model as the initial model in the next round. In distributed federated learning, the central sensor in the star topology also performs the same computation but its model will not be used as the  model by other sensors in the next round. As a consequence, the performance achieved in the star topology cannot be comparable to conventional federated learning.

Table~\ref{table:converge} shows that the accuracy of federated learning for the star topology is the worst one when the results converge. The line topology, although it is good for the accuracy of federated learning, makes federated learning take the largest number of rounds to converge. In general, a network of sensors can help federated learning achieve fast convergence and good accuracy if each sensor has many neighbors. An example is the grid topology, where most sensors have four neighbors. When applied in the grid topology, federated learning converges fast ($581$ rounds) and achieves high accuracy ($0.9389$).

\subsection{Effect of Packet Losses}

Wireless transmissions may fail due to various reasons, e.g., interference, jamming, and dynamic channel effects. Thus, we consider packet losses with certain probability. We assume that the deep neural network model of each sensor is transmitted as a packet. If a packet is lost, the receiver cannot receive the updated model in this packet and thus will not count it in the federated averaging process.  Table~\ref{table:converge-failure} shows convergence results under random packet losses for the network with a random topology. We can see that distributed federated learning is robust to packet losses. There is not much change in accuracy even if the packet loss probability is $50$\%. The convergence time changes with different packet loss probabilities but more packet losses incurred may not always increase the convergence time.


\subsection{Reduction of Communication Overhead and Energy Consumption}
\label{subsec:skip}

So far, we assume that all sensors participate in the update process in all rounds. The communication overhead is $N$ broadcasts, where $N$ is the number of sensors in the network. Each sensor needs to spend its (transmit) energy in all rounds. Next, we study whether we can reduce the communication overhead and energy consumption. One approach is letting a sensor broadcast its update with certain probability in each round. For the network with a random topology, Table~\ref{table:converge-overhead} shows that we can reduce the broadcast probability to $25$\% without decreasing the accuracy. But if we further reduce the broadcast probability to $10$\%, the accuracy drops from $92.21$\% to $88.65$\%. We can see that the convergence time is reduced, since the improvement per round is smaller when the broadcast probability is smaller. These results show that we can reduce the communication overhead and energy consumption by reducing the rate of broadcasts. 

\begin{table}[t!]
	\caption{Performance of federated learning for different packet loss probabilities.}
	\vspace{0.25cm}
	\centering
	{\small
		\begin{tabular}{c|c|c} 
Loss probability & Time & Best average accuracy \\ \hline \hline
$0$\% & $704$ & $0.9221$ \\ \hline
$5$\% & $721$ & $0.9257$ \\ \hline
$20$\% & $696$ & $0.9247$ \\ \hline
$50$\% & $785$ & $0.9302$
		\end{tabular}
	}
	\label{table:converge-failure}
\end{table}
\begin{table}
	\caption{Performance of federated learning for different broadcast probabilities.}
	\vspace{0.25cm}
	\centering
	{\small
		\begin{tabular}{c|c|c} 
Broadcast probability & Time & Best average accuracy \\ \hline \hline
$100$\% & $704$ & $0.9221$ \\ \hline
$50$\% & $505$ & $0.9234$ \\ \hline
$25$\% & $455$ & $0.9110$ \\ \hline
$10$\% & $622$ & $0.8865$
		\end{tabular}
	}
	\label{table:converge-overhead}
\end{table}


\section{Conclusion}
\label{sec:conclusion}

In this paper, we designed a distributed federated learning scheme and applied it to a multi-hop network of sensors for distributed spectrum sensing. Each sensor collects its own I/Q data via spectrum sensing and aims to build a deep learning model with collaboration of other sensors to detect signal types not only for its own data, but also for data collected by other sensors. The conventional federated learning scheme requires a server that can communicate with all sensors (over a single hop), which may not be possible for a multi-hop network. The distributed federated learning scheme removes this requirement by letting each sensor exchange model updates with neighbors. We showed that this scheme works well for various network topologies and in general, a network with more connections among nodes can achieve fast convergence time and high accuracy for all local models. We also showed that this scheme is robust to packet losses. Moreover, sensors may reduce the number of times they broadcast their model updates to reduce communication overhead and energy consumption. We showed that distributed federated learning can still converge to a global model with high accuracy.

\bibliographystyle{spiebib} 

\begin{thebibliography}{99}

\bibitem{McMahan17:FL}
H.~B.~McMahan and D.~Ramage, ``Federated Learning: Collaborative Machine Learning without Centralized Training Data", \emph{Google AI Blog}, April~2017. Available at https://ai.googleblog.com/2017/04/federated-learning-collaborative.html.



\bibitem{Survey0:FL}
P. Kairouz, et al., ``Advances and Open Problems in Federated Learning," \emph{Foundations and Trends in Machine Learning}, 14, no. 1-–2 (2021): 1--210.

\bibitem{Survey1:FL}
K. Bonawitz, et al. ``Towards Federated Learning at Scale: System Design," \emph{Proceedings of Machine Learning and Systems}, 1 (2019): 374--388.

\bibitem{Survey2:FL}
Q. Yang, Y. Liu, T. Chen, and Y. Tong, ``Federated Machine Learning: Concept and Applications," \emph{ACM Transactions on Intelligent Systems and Technology (TIST)}, vol 10, no. 2, March~2019.


\bibitem{Survey3:FL}
T. Li, A. K. Sahu, A. Talwalkar, and V. Smith, ``Federated Learning: Challenges, Methods, and Future Directions," \emph{IEEE Signal Processing Magazine}, vol. 37, no. 3, pp. 50--60, May 2020.

\bibitem{Wireless5:FL}
N. H. Tran, W. Bao, A. Zomaya, M. N. H. Nguyen, and C. S. Hong, ``Federated Learning over Wireless Networks: Optimization Model Design and Analysis," \emph{IEEE INFOCOM}, pp. 1387--1395, 2019.

\bibitem{Wireless4:FL}	
S. Niknam, H. S. Dhillon, and J. H. Reed, ``Federated Learning for Wireless Communications: Motivation, Opportunities, and Challenges," \emph{IEEE Communications Magazine}, vol. 58, no. 6, pp. 46--51, June 2020.

\bibitem{Wireless1:FL}
M. Chen, Z. Yang, W. Saad, C. Yin, H. V. Poor, and S. Cui, ``A Joint Learning and Communications Framework for Federated Learning Over Wireless Networks," \emph{IEEE Transactions on Wireless Communications}, vol. 20, no. 1, pp. 269--283, Jan. 2021.

\bibitem{Wireless2:FL}	
Z. Yang, M. Chen, W. Saad, C. S. Hong, and M. Shikh-Bahaei, ``Energy Efficient Federated Learning Over Wireless Communication Networks," \emph{IEEE Transactions on Wireless Communications}, vol. 20, no. 3, pp. 1935--1949, March~2021.

\bibitem{Wireless3:FL}	
M. Chen, D. Gunduz, K. Huang, W. Saad, M. Bennis, A. V. Feljan, and H. V. Poor, ``Distributed Learning in Wireless Networks: Recent Progress and Future Challenges," \emph{IEEE Journal on Selected Areas in Communications}, vol. 39, no. 12, pp. 3579--3605, Dec. 2021.





\bibitem{MEC:FL}
W. Y. B. Lim, N. C. Luong, D. T. Hoang, Y. Jiao, Y. -C. Liang, Q. Yang, D. Niyato, and C. Miao, ``Federated Learning in Mobile Edge Networks: A Comprehensive Survey," \emph{IEEE Communications Surveys \& Tutorials}, vol. 22, no. 3, pp. 2031--2063, 2020.

\bibitem{FL:IoT}
L. U. Khan, W. Saad, Z. Han, E. Hossain, and C. S. Hong, ``Federated Learning for Internet of Things: Recent Advances, Taxonomy, and Open Challenges," \emph{IEEE Communications Surveys \& Tutorials}, vol. 23, no. 3, pp. 1759--1799, third quarter 2021.

\bibitem{5G:FL}
Y. Liu, J. Peng, J. Kang, A. M. Iliyasu, D. Niyato and A. A. A. El-Latif, ``A Secure Federated Learning Framework for 5G Networks," \emph{IEEE Wireless Communications}, vol. 27, no. 4, pp. 24--31, Aug. 2020,

\bibitem{6G:FL}
Z. Yang, M. Chen, K. Wong, H. V. Poor, and S. Cui, ``Federated Learning for 6G: Applications, Challenges, and Opportunities," \emph{Engineering}, 2021.

\bibitem{Multi1:FL}
C.~He, C.~Tan, H.~Tang, S.~Qiu, and J.~Liu, ``Central Server Free Federated Learning over Single-sided Trust Social Networks," arXiv preprint arXiv:1910.04956 (2019).

\bibitem{Multi2:FL}
A.~Lalitha, O.~C.~Kilinc, T.~Javidi, and F.~Koushanfar, ``Peer-to-peer Federated Learning on Graphs," arXiv preprint arXiv:1901.11173 (2019).

\bibitem{Attack:FL}
Y.~Shi and Y.~E.~Sagduyu, ``Jamming Attacks on Federated Learning in Wireless Networks," arViv preprint arXiv:2201.05172, 2022.

\bibitem{FCC}
Code of Federal Regulations, ``Citizens Broadband Radio Service," Title 47, Part 96, 2015.

\bibitem{DL:CBRS}
W. M. Lees, A. Wunderlich, P. J. Jeavons, P. D. Hale, and M. R. Souryal, ``Deep Learning Classification of 3.5-GHz Band Spectrograms with Applications to Spectrum Sensing," \emph{IEEE Transactions on Cognitive Communications and Networking}, vol. 5, no. 2, pp. 224–-236, 2019.

\bibitem{DL:wireless1}
Y. Shi, K. Davaslioglu, Y. E. Sagduyu, W. C. Headley, M. Fowler, and G. Green, ``Deep Learning for Signal Classification in Unknown and Dynamic Spectrum Environments," \emph{IEEE International Symposium on Dynamic Spectrum Access Networks (DySPAN), 2019}.

\bibitem{DL:wireless2}
T. Erpek, T. O'Shea, Y. E. Sagduyu, Y. Shi, and T. C. Clancy, ``Deep Learning for Wireless Communications," in Development and Analysis of Deep Learning Architectures, Springer, 2020.



\end{thebibliography}

\end{document}